\documentclass[aps,prl,twocolumn,superscriptaddress,floatfix,showpacs]{revtex4}

\usepackage{amsmath,amssymb}
\usepackage{graphicx}
\begin{document}                              
\title{Minimal Single-Particle Hamiltonian for Charge Carriers
in Epitaxial Graphene on 4{\it H}-SiC(0001)}
\author{Seungchul Kim}
\altaffiliation
{Present Address: The Makineni Theoretical Laboratories,
Department of Chemistry, 
University of Pennsylvania, Philadelphia, PA 19104, USA}
\affiliation
{Department of Physics and Astronomy, 
Seoul National University, Seoul 151-747, Korea}
\author{Jisoon Ihm}
\affiliation
{Department of Physics and Astronomy, 
Seoul National University, Seoul 151-747, Korea}
\author{Hyoung Joon Choi}
\affiliation
{Department of Physics and IPAP, 
Yonsei University, Seoul 120-749, Korea}
\author{Young-Woo Son}
\email{hand@kias.re.kr}
\affiliation
{Korea Institute for Advanced Study, Seoul 130-722, Korea}
\date{\today}
\begin{abstract}
We present a minimal but crucial microscopic theory for 
epitaxial graphene and graphene nanoribbons on the 4$H$-SiC(0001) 
surface -- protopypical materials to explore physical properties of graphene in large scale.
Coarse-grained model Hamiltonians are constructed based on the atomic and
electronic structures of the systems from first-principles calculations.
From the theory, we unambiguously uncover origins of several intriguing experimental observations
such as broken-symmetry states around the Dirac points and 
new energy bands arising throughout the Brillouin zone, thereby 
establishing the role of substates in modifying electronic properties of graphene.
We also predict that armchair graphene nanoribbons 
on the surface have a single energy gap of 0.2 eV 
when their widths are over 15 nm, in sharp contrast to
their usual family behavior.
\end{abstract}
%73.20.-r 	Electron states at surfaces and interfaces
%81.05.Uw 	Carbon, diamond, graphite
%68.35.-p 	Solid surfaces and solid¿solid interfaces: structure and energetics
%71.20.-b 	Electron density of states and band structure of crystalline solids

\maketitle                            %APS format
Graphene has attracted immense interests
because of the unusual relativistic energy dispersions 
with the chiral massless Dirac fermions 
near the Fermi level ($E_F$)~\cite{novoselov0,novoselov,zhang}. 
The direct observation of such peculiar quasiparticle spectra in graphene 
is particularly important not only for understanding 
its novel physical properties~\cite{novoselov,zhang,berger,geim} 
but also for practical applications~\cite{berger,geim}. 
So, several measurements through the high resolution 
angle resolved photoemission spectroscopy (ARPES) 
are performed on a single layered epitaxial graphene lying 
on the silicon carbide (0001) surfaces~\cite{bostwick,zhou,
rotenberg,zhou2,lee}. 
Surprisingly, the reported quasiparticle spectra reveal 
anomalous energy dispersions around the Dirac energy point ($E_D$)
indicating highly renormalized bands~\cite{bostwick} 
or energy gap~\cite{zhou} there. 
Many experimental~\cite{rotenberg,zhou2,lee,mallet,ohta,
bostwick2,vitali,brihuega,hass,ohta2,lauffer} and 
theoretical studies~\cite{mucha,mattausch,
varchon,kim,varchon2,benfatto,trevis,park}
address these important problems 
but there is still no consensus on origins of the anomalous
spectrum.
Moreover, there is no clear understanding
on other anomalous ARPES observations such as
the broken six-fold symmetry near the $E_D$
and new distorted hexagonal energy bands around 
Brillouine zone (BZ) center ~\cite{bostwick,zhou,rotenberg,
zhou2}. 

The complex interfacial structures arise 
when epitaxial graphene is grown by annealing SiC 
surfaces~\cite{tromp}. 
During the thermal decomposition of the surfaces, 
a layer of carbon atoms, called the buffer layer, 
forms first without exhibiting the typical linear energy 
dispersion of $\pi$-states near the $E_F$~\cite{tromp,emtsev,heer,emtsev2}. 
Then, on top of the buffer layer, the clean honeycomb lattice 
of carbon atoms grows~\cite{tromp,emtsev,heer,emtsev2,chen}. 
The lattice mismatch between the SiC(0001) surface, buffer layer and
graphene gives rise to the large scale surface reconstruction 
with a periodicity of $6\sqrt{3}\times6\sqrt{3}R30^\circ$
(in short, $6R3$) with respect to the SiC(0001) 
surface unitcell, which is observed 
in the low energy electron diffraction (LEED) 
measurements~\cite{tromp,emtsev,heer,emtsev2,chen}.
The scanning tunnelling microscopy (STM) 
image, however, indicates an approximate $6\times6$ 
periodicity~\cite{mallet,vitali,brihuega,hass} of the reconstruction.  
We note that such superperiodic rearrangements shall 
impose constraints on possible theoretical models 
to explain the anomalous electronic structures mentioned above.

In this paper, we show that the interactions 
between epitaxial graphene and the reconstructed layer underneath it 
are the main driving forces to several anomalous features 
observed in recent experiments~\cite{bostwick,
zhou,lee,rotenberg,zhou2,mallet,ohta,bostwick2,vitali,
brihuega,hass,ohta2,lauffer} 
on electronic properties of epitaxial graphene. 
From the simulated ARPES spectra on the systems, 
it is shown that the symmetry breaking at the $E_D$ 
and the new hexagonal bands throughout 
the two-dimensional BZ have the same origins. 
Based on the model, we predict that graphene nanoribbons
or finite size fragments of graphene~\cite{emtsev,zhou2,ohta,berger}
show an homogeneous energy gap contrary to their 
family behavior~\cite{son}.
Moreover, by extending our microscopic model 
for monolayer epitaxial graphene to bilayer one, 
we identify the effects of the buffer layer to its energy spectrum 
and characteristic energy gap~\cite{zhou,lee,rotenberg,zhou2,ohta2}. 
Our computational results indicate that the interplays 
between geometries and electronic structures are pivotal 
in altering global energy bands of graphene, 
notwithstanding that the many-body interactions~\cite{benfatto,
trevis,park} are expected to play some roles 
in modifying the quasi-particle spectrum 
near the $E_D$ locally.

From the {\it ab-initio} pseudopotential density 
functional method~\cite{kim,soler,epaps},
we find that the carbon atoms in the buffer layer 
with superperiodic $6R3$ unitcell are split 
into lattice matched regions, where carbon atoms 
have $\sigma$-bonds to silicon atoms of the 4$H$-SiC(0001) surface, 
and their boundaries without the $\sigma$-bonds~\cite{kim} (Fig. 1(a)). 
The carbon atoms at the boundaries of the lattice matched 
regions exhibit an approximate $6\times6$ domain satisfying 
the geometric constriction imposed simultaneously 
by both LEED and STM measurements.
To explore various initial conditions for buffer layer formation, 
we shift the initial atomic coordinates of the buffer layer (shown in Fig. 1(b))
on top of the 4$H$-SiC(0001) surface by either
$\frac{1}{2}{\bf a}_1 $ (Fig. 1(c)) or $\frac{1}{2}{\bf a}_2$ (Fig. 1(d))
where ${\bf a}_{1(2)}$ is an unit vector of graphene. 
We find that the final relaxed atomic geometries for all initial coordinates 
are essentially same to each other except for minor differences 
in the connectivity of $\pi$-electrons along quasi-$6\times6$ 
domain boundaries (Figs. 1(b)-(d)). 

We build up a minimal (coarse-grained) microscopic model 
for interactions between epitaxial graphene 
and the buffer layer based on atomic and 
electronic structures obtained from our first-principles calculation 
(See detailed method in~\cite{epaps}). 
From the first-principles calculations,
it is found that 
the interactions between $\pi$-orbital states at the domain boundary 
of the buffer and ones in graphene play the most significant role 
to determine the electronic structures of the system
while the states of atoms inside the domain and those under the buffer layer 
have negligible contributions to the electronic structures
near the $E_F$ and $E_D$. 
Hence, it is sufficient to approximate the buffer layer 
to the coarse-grained atomic configuration of the quasi-$6\times6$ periodic 
connections of $\pi$-electrons only (thick black lines in Figs. 1(b)-(d)). 
Our tight-binding Hamiltonian for monolayer epitaxial graphene 
on coarse-grained buffer layer model (Fig. 1b) can be written as
\begin{eqnarray}
{\mathcal H} &=& -t_{\rm G}\sum_{\langle i,j\rangle}c^\dagger_{i}c_{j}
			       -V_{\rm G}\sum_{i}c^\dagger_{i}c_{i} 
			    -t_{\rm B}\sum_{\langle l,m\rangle}b^\dagger_l b_m \nonumber\\
			& &	   -V_{\rm B}\sum_{l} b^\dagger_l b_l 
		    	-\gamma \sum_{\langle i,m\rangle}c^\dagger_{i} b_m +\rm{(c.c.)},
\end{eqnarray}
where $t_{\rm G}$ (2.70 eV) and $t_{\rm B}$ (1.50 eV) are 
the nearest neighbour hopping amplitude between carbon atoms in graphene 
and those in the buffer layer respectively. 
The $c_i$ and $b_l$ are annihilation operators for electron in graphene 
and buffer layer respectively. 
$\gamma$ (0.30 eV) denotes the interlayer hopping amplitude 
between the nearest neighbour carbon atoms belong to graphene and 
the buffer layer with the Bernal type stacking respectively. 
$V_{\rm G}$ (0.35 eV) and $V_{\rm B}$ (0.34 eV) describe 
the potential for graphene and the buffer layer considering 
charge redistributions due to the polar SiC surface. 
Since our model Hamiltonians are described 
within the single-orbital tight-binding approximation, 
we can readily extend our model to explore 
the electronic structures of finite-sized monolayer 
and multilayer epitaxial graphene respectively 
which are beyond the reach of the first-principles calculations.

\begin{figure}[t]
\includegraphics[width=8.5 cm]{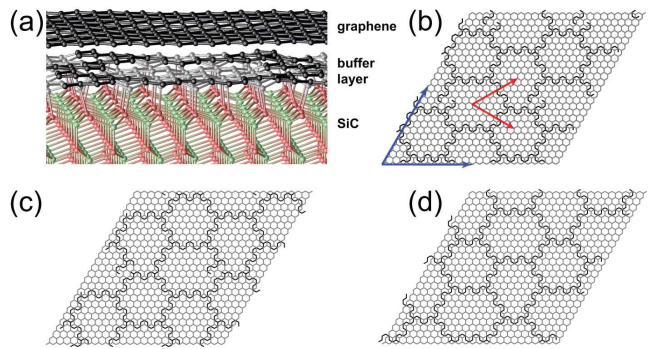}
\caption{(color online) (a) Ball and stick model for fully relaxed
atomic configuration of epitaxial graphene, buffer layer 
and 4$H$-SiC(0001) surface with $6R3$ supercell. 
Epitaxial graphene (black) is located 
on top of the buffer carbon atoms (black and grey). 
The buffer carbon atoms consisting $6\times6$ domain
are denoted by grey color and its boundary by black.
The silicon and carbon atoms in the SiC are denoted by 
red and green respectively. 
(b)-(d) Coarse-grained minimal microscopic models. 
The red thick arrows indicate the $6\times6$ supercell unit vectors 
and the blue ones $6R3$ unit vectors. 
Black lines represent the connectivity of $\pi$-electrons 
of the buffer carbon atoms with approximate $6\times6$ 
domain boundaries and grey lines denote 
the hexagonal network of epitaxial graphene 
on top of the buffer. 
}
\end{figure}

We have found that the nearest neighbour inter- and 
intra-layer interactions between $\pi$-electrons 
in the coarse-grained atomic model (Eq. (1)) 
are sufficient to reproduce the electronic energy bands 
obtained from the first-principles calculations  (Fig. 2)
(for detailed comparisons, see~\cite{epaps}). 
We notice that inclusion of the nearest neighbour 
interlayer interaction between graphene and 
the coarse-grained buffer layer model already 
breaks the symmetry between two sublattices of graphene. 
The simulated ARPES spectrum for monolayer epitaxial graphene 
agrees well with both our previous results~\cite{kim} 
from the first-principles calculation and 
ones~\cite{bostwick,zhou,lee,rotenberg,zhou2} 
from experiments (Fig. 2) 
(for detailed ARPES simulation method, see~\cite{epaps}). 
When the initial conditions for the buffer layer formation
are varied and the resulting geometries
for the approximated $6\times6$ domain are slightly altered as
shown in Figs. 1(b)-(d),
the simulated ARPES spectrum display essentially
same structures with energy gaps at the $E_D$'s and
the midgap states as shown 
in Figs 2(a)-(c).  
Hence, We confirm from the simple microscopic model
that the sublattice symmetry-breaking 
interlayer interaction indeed opens a gap of $\sim$0.20 eV 
at the $E_D$ and 
the presence of midgap states give rise 
to high ARPES intensities inside the energy gap  
as well as the level repulsion 
between upper and lower Dirac cones (Fig. 2(d)). 

\begin{figure}[t]
\includegraphics[width= 8.5 cm]{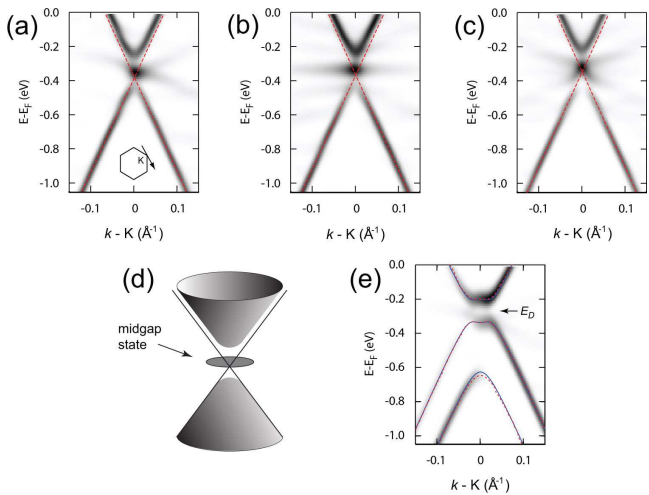}
\caption{(color online)
Simulated ARPES spectrum near the $E_D$  
for various atomic models are drawn 
along the arrow shown in the inset of (a). 
The superimposed dotted lines are 
an ideal energy spectrum of graphene. 
Each simulated ARPES spectra in (a)-(c) corresponds 
to the geometry shown in Fig. 1(b)-(d), respectively. 
The positions of $E_D$ and size of energy gaps are 
(a) $-0.35$ eV, 0.20 eV, (b) $-0.33$ eV, 0.19 eV, and 
(c) $-0.33$ eV, 0.23 eV , respectively.
(d) Schematic energy dispersion for epitaxial graphene 
with gap and midgap state. 
The straight lines are linear dispersion relations 
of ideal graphene. 
(e) Simulated ARPES spectra of bilayer epitaxial graphene
near the $E_D$. Superimposed red and 
blue lines represent energy bands of bilayer graphene
with and without buffer layer respectively.
The position of $E_D$ and energy gap are given by
-0.26 eV and  0.12 eV respectively.
}
\end{figure}

From simulated ARPES spectrum for bilayer epitaxial graphene, 
we find that the $E_D$ approaches
to the $E_F$ and the energy gap decreases 
from 0.20 to 0.12 eV (Fig. 2(e)), 
agreeing well with experimental observations~\cite{zhou,
lee,ohta2} 
Due to the charge transfer between the SiC surfaces and graphene, 
perpendicular electric fields exist 
on the multilayer epitaxial graphene opening an energy 
gap~\cite{mccann,ohta3}. 
Though the fingerprints of midgap states are hardly visible
for the bilayer epitaxial graphene,  
it is shown that the interaction 
between the buffer and graphene flattens 
the characteristic Mexican-hat-shaped band~\cite{mccann} 
at the bottom of the upper Dirac cone [Fig. 2(e)and Fig. 1S (c)
in~\cite{epaps}] and
shifts the second subband in the lower cone 
downward by 20 meV (Fig. 2(e)). 

\begin{figure}[t]
\includegraphics[width= 8.0 cm]{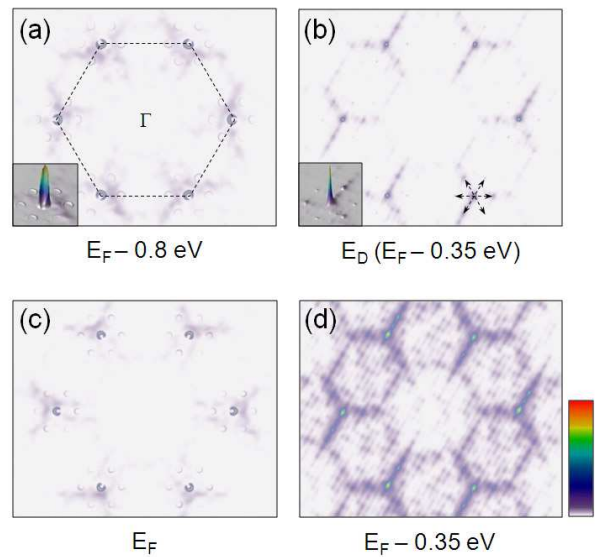}
\caption{(color online)
Simulated ARPES intensity maps for constant energy taken 
on monolayer epitaxial graphene at (a) $-$0.8 eV 
(b) $-$0.35 eV ($E_D$) and (c) 0.0 eV ($E_F$) 
respectively. 
The first Brillouine zone of graphene is 
denoted by dotted line in (a). 
The insets in (a) and (b) show the stereographic plots 
near the $E_D$. 
The small arrows in (b) indicate the second shortest 
reciprocal lattice vectors of $6R3$
supercell, which connect the $K$-point to six faint replicas. 
(d) Simulated ARPES intensity maps taken 
on buffer layer at $-$0.35 eV ($E_D$) without graphene. 
The bar on the right side denotes 
the color scale of relative intensities 
from zero (white) to maximum (red).
}
\end{figure}

By simulating the two-dimensional ARPES intensity 
maps with various fixed energies (Fig. 3), 
we find that the interactions between the buffer layer 
and graphene indeed induce the symmetry breaking 
near the $E_D$ and produce new hexagonal energy bands 
throughout the first BZ of graphene. 
Our simulated fixed-energy intensity patterns 
give a good agreement with experimental 
results~\cite{bostwick,zhou,rotenberg,zhou2} (Fig. 3). 
Due to the two-source interference between photo-excited electrons 
from two equivalent atomic sites ({\it \`a la} Young's double slit) 
or helical nature of the carriers in graphene, 
the intensity patterns at the $K$-points show 
the typical crescent shape anisotropy~\cite{mucha}. 
With the characteristic high intensity at the $K$-points, 
there exist six equivalent faint replicas around each $K$-point 
forming a smaller hexagon (inset in Fig. 3(a)). 
When approaching $E_D$, the highest intensities at the $K$-points 
become isotropic and three of six replicas become weaker compared 
to other three points (inset in Fig. 3(b)). 
The intensities of stronger three satellites amount to 7\% 
of the main peak at the $K$-points while those of the weaker to 2\% 
agreeing with experiment results qualitatively~\cite{zhou,
rotenberg,zhou2}. 
The replicas are connected by the second-shortest reciprocal 
lattice vectors of the $6R3$ supercell (Fig. 3(b)) 
and the area of the smaller hexagon nearby is 
$\frac{3}{13}\times\frac{3}{13}$
of the first BZ of graphene~\cite{zhou}. 
When the interlayer interaction is set to zero intentionally 
in our simulation, we cannot find any symmetry breaking phenomena. 
Thus, we conclude that apparent six-fold symmetry breaking 
near the $K$-points is due to the interaction between the buffer 
and graphene. 
Together with six faint replicas, 
there are global faint features throughout the first BZ (Fig. 3(a)-(c)). 
We also find that the larger hexagonal structure around 
the $\Gamma$-point observed 
in the experiments~\cite{zhou,rotenberg,zhou2} (25\% of total area) 
originates from the buffer layer. 
By simulating constant energy map (Fig. 3(d)) and ARPES spectrum (Fig. 2S in~\cite{epaps}) 
of our coarse-grained buffer layer model, 
we show that the underlying faint features are 
reminiscent of the interlayer interactions 
between the buffer and graphene.

\begin{figure}[t]
\includegraphics[width= 8.5 cm]{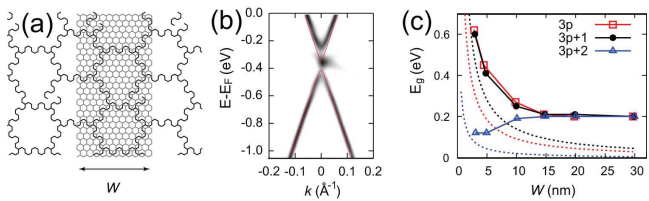}
\caption{(color online)
(a) Atomic model for EAGNR of width $W$  
(here, 24-EANGR is drawn)
on top of the buffer layer. 
The lines follow the convention introduced in Fig. 1. 
(b) Simulated ARPES spectrum for nanoribbons 
with $W=19.9$ nm (163-EAGNR). 
The superimposed lines represent the energy 
band of 163-EAGNR having an energy gap of 0.72 eV 
without the buffer. 
(c) The energy gaps of $N$-EAGNRs as a function of width
($N = 3p, 3p+1, 3p+2, p=$ positive integer). 
The dotted lines correspond to energy energy gaps 
belong to each family without buffer layer.
}
\end{figure}

Usually, the surface of epitaxial graphene exhibits 
the finite-sized terrace patterns~\cite{emtsev,rotenberg,ohta}.
Thus the quantum confinement effect~\cite{son} may play a role 
in determining the energy gap~\cite{zhou2}. 
To illustrate such an effect, we calculate 
electronic structures of epitaxial graphene nanoribbons 
with armchair edges on both sides (in short, EAGNR) on top 
of the buffer layer. 
Following the convention~\cite{son}, three families of EAGNRs are 
denoted by the number of dimer lines, $N$, i.e., $N$-EAGNR 
($N=3p$, $3p+1$, $3p+2$ families, $p$ is a positive integer). 
By using the same model for the single layer epitaxial 
graphene (Fig. 4(a)), we show that the EAGNRs have 
a similar ARPES spectrum to epitaxial graphene for energy gaps, 
midgap states, and level repulsion between upper 
and lower cones, respectively (Fig. 4(b)). 
Moreover, when the width of EAGNR is over 15 nm, 
the typical family behaviour of energy gaps~\cite{son} disappears 
completely and converges to the energy gap (0.20 eV) 
of two-dimensional epitaxial graphene (Fig. 4(c)). 
Hence, we can conclude that the terrace patterns or finite-sized
epitaxial graphene exhibit essentially the same electronic
structures of the ideal two dimensional one. 
On the other hand, the present calculation results
indicate that epitaxially grown graphene
nanoribbons on the SiC surface will have the
homogenous energy gap if the width is over 15 nm.

In summary, we have constructed the microscopic theory
for expitaxial graphene on 4$H$-SiC(0001) surface
incorporating interactions between graphene and the surface.
The simulated experimental observations
based on the theory have been shown to explain
the several atypical aspects of epitaxial graphene
from a single and unified view and, thus, shed
light on understanding the quasiparticle spectrum
of graphene in various circumstances.

S. K and J. I acknowledge the support of the KOSEF 
through the SRC program 
(Center for Nanotubes and Nanostructured Composites). 
H. J. C. was supported by the KRF (KRF-2007-314-C00075) 
and by the KOSEF Grant No. R01-2007-000-20922-0. 
Y.-W. S. was supported by Quantum Metamaterials 
Research Center No. R11-2008-053-01002-0 
and Nano R\&D program 2008-03670 
through the KOSEF funded 
by the Korean government (MEST). 
Computational resources have been provided 
by KISTI (KSC-2008-S02-0004) and the KIAS Linux Cluster System.

\section{SUPPLEMENTARY INFORMATION}

\subsection{Construction of minimal model Hamiltonian within tight-binding approximations}
In first-principles calculations, 
we expand the wave function with localized basis sets~\cite{soler,kim}
to handle a large number of atoms in the system (typically $>$ 1600 atoms). 
A single-$\zeta$ for hydrogen, a singe-$\zeta$ plus polarization for silicon
and a mixed basis set with a single and double-$\zeta$ 
for $s$- and $p$-orbitals of carbon atom have been used, respectively~\cite{kim}. 
The Kleinman-Bylander's fully separable nonlocal
projectors~\cite{kb} are used in the norm-conserving 
pseudopotentials~\cite{psp} and the local density approximation~\cite{alder} is
employed in setting up the exchange-correlation potential.
We have thoroughly tested our basis set with other calculation parameters 
to reproduce the atomic and electronic structures of SiC, graphene, 
other model for epitaxial graphene studied 
in previous literatures~\cite{mattausch,varchon,varchon2}, respectively. 
We modeled the 4$H$-SiC(0001) substrate in the simulation 
with four alternating silicon and carbon atomic layers. 
Hydrogen atoms are introduced to passivate the dangling bonds 
in bottom of the slab.
On top of the Si-terminated surface of 4$H$-SiC(0001), 
one, two, and three graphene layers 
are placed for the buffer layer, monolayer graphene, and 
bilayer graphene, respectively. 

Based on the low energy electron diffraction 
(LEED) experiments~\cite{berger,heer,chen}, 
the large supercell with the $6\sqrt{3}\times6\sqrt{3}R30^\circ$ (in short $6R3$) 
periodicity (equivalent to $13\times13$ times graphene unit cell) 
is imposed to the calculations. 
The atomic positions are determined by total energy minimization 
calculations until the forces on each atom are less than 0.06 eV/\AA~while 
atoms belonging to the last two silicon and carbon layers are fixed 
to the bulk atomic structure of 4$H$-SiC. 
In the electronic structure calculations after geometric optimization process, 
we use $2\times2$ k-point sampling in lateral directions and 
set very large size of vacuum (50 \AA) in surface normal direction 
to prevent the spurious dipolar interactions 
between terminated slab geometries in supercell configuration. 

\begin{figure}[b]
\includegraphics[width=8.5cm]{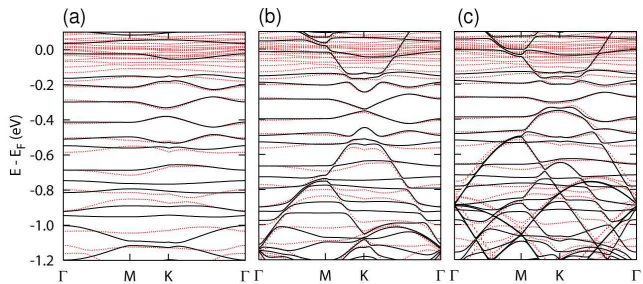}
\caption[Fig. 1S]{
Comparison between electronic energy bands 
from the first-principles calculation and tight-binding approximation. 
Electronic energy bands of (a) the buffer layer 
on top of 4$H$-SiC(0001) slab model,
(b) monolayer epitaxial graphene and (c) bilayer epitaxial graphene 
on top of the buffer layer with 4$H$-SiC(0001) slab, respectively. 
The spectrum are drawn along the high symmetric lines of 
the first Brillouine zone of $6R3$ supercell. 
The dotted red lines are obtained by the first-principles calculations 
on the atomic models including one (two) graphene layer, 
buffer layer, and 4$H$-SiC(0001) slab. 
The solid black lines are obtained by our minimal microscopic model 
Hamiltonian within the tight-binding approximations (Eq. 1). 
It is found that the dense flat bands near the Fermi levels are 
from the states isolated inside the approximate $6\times6$ domains 
of the buffer layer and SiC surface underneath it, 
which do not contribute to ARPES intensities. 
}
\end{figure}

During the high temperature thermal decomposition process~\cite{berger,heer,chen,ohta2,bostwick,zhou,lauffer}, 
the detailed atomic structures and shape of the domain 
may vary from sample to sample. 
However, the domain boundary of $\pi$-orbitals of the buffer carbon atoms 
and the exact $6R3$ periodicity 
remains the same as observed in many experiments~\cite{berger,heer,chen,ohta2,bostwick,zhou,lauffer}
and shown in our calculations~\cite{kim}. 
Hence, it is sufficient to approximate the buffer layer 
to the coarse-grained atomic configuration of the quasi-$6\times6$ periodic 
connections of $\pi$-electrons only. 
We also found from the first-principles calculations 
that the interactions between $\pi$-orbital states at the domain boundary 
of the buffer and ones in graphene play the most significant role 
to determine the electronic structures of the system~\cite{kim}. 
It is also noticeable that electronic states of atoms 
inside the domain and those under the buffer layer have negligible contributions. 
\begin{figure*}[t]
\includegraphics[width=16cm]{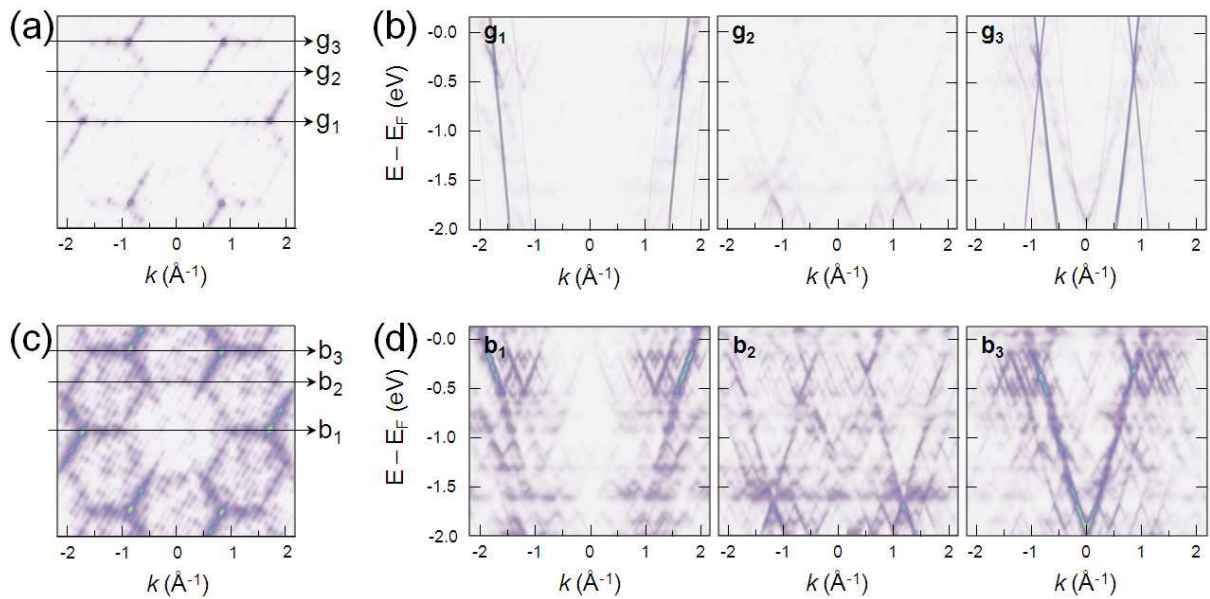}
\caption{
Simulated energy spectrum of epitaxial graphene 
and buffer layer along various directions in the first Brillouine 
zone of graphene. 
(a) Simulated ARPES intensity map taken on monolayer 
epitaxial graphene at the Dirac energy point ($E_D$). 
(b) From left to right panels, simulated ARPES spectrum 
along the $g_1$, $g_2$ and $g_3$ shown in (a) respectively. 
Together with strong ARPES intensities from graphene, 
there are several dispersions originating from the buffer layer. 
(c) Simulated ARPES intensity map taken on minimal buffer 
layer model at the $E_D$.
(d) From left to right panels, simulated ARPES spectrum 
along the $b_1$, $b_2$ and $b_3$ shown in (c) respectively. 
It is noticeable that all characteristic ARPES spectrum 
of the buffer shown in each panel of (d) appears faintly 
in the corresponding one of (b), respectively.
}
\end{figure*}

Hence, based on results from the first-principles calculations, 
our tight-binding Hamiltonian for monolayer epitaxial graphene 
on coarse-grained buffer layer model (Fig. 1(b)-(d)) can be written as
\begin{eqnarray}
{\mathcal H} &=& -t_{\rm G}\sum_{\langle i,j\rangle}c^\dagger_{i}c_{j}
			       -V_{\rm G}\sum_{i}c^\dagger_{i}c_{i} 
			    -t_{\rm B}\sum_{\langle l,m\rangle}b^\dagger_l b_m \nonumber\\
			& &	   -V_{\rm B}\sum_{l} b^\dagger_l b_l 
		    	-\gamma \sum_{\langle i,m\rangle}c^\dagger_{i} b_m +\rm{(c.c.)},
\end{eqnarray}
where $t_{\rm G}(= 2.70 \rm{eV})$ and $t_{\rm B}(= 1.50 \rm{eV})$ are 
the nearest neighbour hopping amplitude between carbon atoms in graphene 
and those in the buffer layer respectively. 
The $c_i$ and $b_l$ are annihilation operators for electron in graphene 
and buffer layer respectively. 
$\gamma(= 0.30 \rm{eV})$ denotes the interlayer hopping amplitude 
between the nearest neighbour carbon atoms belong to graphene and 
the buffer layer with the Bernal type stacking respectively. 
$V_{\rm G} (= 0.35 \rm{eV})$ and $V_{\rm B} (= 0.34 \rm{eV})$ describe 
the potential for graphene and the buffer layer considering 
charge redistributions due to the polar SiC surface. 
These parameters were found to be enough for fitting our 
first-principles energy bands of the buffer layer and
monolayer epitaxial graphene, respectively (Fig. 1S (a) and (b)) 
while the second and third nearest neighbour interlayer interaction 
terms improve the agreements a little. 

For bilayer epitaxial graphene, we introduce another interlayer 
hopping amplitude $\lambda$ between the nearest neighbours 
carbon atoms belong to each graphene layer (Bernal type stacking) 
and potential shifts for each graphene layer, $V_{\rm G}$ and $V_{\rm G}'$ 
with respect to the Fermi level. 
The Hamiltonian for the interaction is written as 
${\mathcal H}'={\mathcal H}_0+{\mathcal H}_1$,
where
\begin{equation}
{\mathcal H}_1=-t_{\rm G}\sum_{\langle i,j\rangle}d^\dagger_{i}d_{j}
			       -V_{\rm G}' \sum_{i}d^\dagger_{i}d_{i} 
	   -\lambda \sum_{i,j}c^{\dagger}_i d_j +\rm{(c.c.)}.
\end{equation}
Here, $d_i$ is the annihilation operator for electron in second graphene. 
We fit the energy spectrum of bilayer epitaxial graphene obtained 
by our model Hamiltonian to the first-principles calculation 
results (Fig. 1S (c)) and found $\lambda = 0.35 \rm{eV}$ 
being quite similar to one of graphite, 
but smaller than 0.48 eV~\cite{ohta2} and 0.46 eV ~\cite{lauffer} 
obtained in recent experiments. 

As shown in Fig. 1S, the agreements between energy spectrums obtained 
by the first-principles calculations and ones by our model Hamiltonians 
are excellent. 
The dense flat bands near the Fermi level in Fig. 1S 
from the first-principles calculations are 
found to originate from localized states inside approximate  $6\times6$ 
domains of the buffer layer and localized states in 4$H$-SiC(0001) 
surfaces underneath the buffer. 
We found that such localized states forming flat bands are 
buried deep inside the surfaces 
and do not contribute to the simulated ARPES spectrum.

\subsection{Simulation of angle resolved photoemission spectroscopy (ARPES) intensities}

We use the Fermi golden rule to simulate ARPES spectra~\cite{shen,mucha}. 
The transition probability ($I$) from an initial Bloch state ($\Psi_i$) 
to an outgoing electron state ($\Psi_f=e^{i{\bf p}\cdot{\bf r}}$) is written as
\begin{equation}
I\sim |\langle\Psi_f |{\mathcal H}_{\rm ph-el}|\Psi_i\rangle|^2 \delta(\hbar\omega+E_i-E_f)
\end{equation}
where ${\mathcal H}_{\rm ph-el}$ is the photon-electron interaction Hamiltonian,
$\omega$ is the frequency of incident photon, and $E_{i(f)}$ 
is an energy of $\Psi_{i(f)}$. 
Tight-binding wavefunction obtained for the system is given by, 
\begin{equation}
\Psi_i =
\sum_j a_j({\bf k})\left[\frac{1}{\sqrt{N}}
              \sum_{\bf T}\phi_j({\bf r}-{\bf x}_j-{\bf T})e^{i{\bf k}\cdot{\bf T}} \right]
\end{equation}
where $a_j ({\bf k})$ is an amplitude for 
the $\pi$-orbital located at ${\bf x}_j$ in the 
$6\sqrt{3}\times6\sqrt{3}R30^\circ$
supercell with unit vector ${\bf T}$ and 
${\bf k}$ is the crystal momentum of electron.

By using the dipole approximation, 
${\mathcal H}_{\rm ph-el}\simeq {\bf A}e^{i{\bf q}\cdot{\bf x}}$, 
the transition probability will be expressed as
\begin{equation}
I({\bf p})\sim \left|\phi_{\bf p}\sum_i 
a_i({\bf k})e^{-i{\bf p}\cdot{\bf x}_i} \right|^2\delta(\hbar\omega+E_i-E_f)
\end{equation}
where ${\bf p}$ is the momentum of outgoing electron and
$\phi_{\bf p}$ is the Fourier transformation of atomic orbital $\phi({\bf r})$, 
defined by $\phi_{\bf p}=\int \phi_j ({\bf r})e^{-i{\bf p}\cdot{\bf r}}d{\bf r}$ 
(set by constant). 
The crystal momentum ${\bf k}$ belong to the first Brillouin zone 
of the $6\sqrt{3}\times 6\sqrt{3}R30^\circ$ supercell 
and satisfys momentum conservation condition in the surface parallel direction 
(i.e., ${\bf k}+{\bf G}={\bf p}_{||}$ for the reciprocal vector $\bf G$) of the supercell. 
We include the attenuation factor for the ARPES intensity considering photoelectron 
mean free path ($\sim$5\AA) to the surface normal direction. 
The conservation of energy imposed by the $\delta$-function 
is replaced by Lorenzian function, 
$\frac{1}{\pi}\frac{\Gamma}{(\hbar\omega+E_i-E_f)^2+\Gamma^2}$, 
with a broadening ($\Gamma$) of 30 meV.  
We simulate the ARPES  spectrum (shown in Figs. 2 and 4(b)) 
with a photon energy of 50 eV 
and the two-dimensional constant energy maps (shown in Fig. 3 and Fig. 2S) 
with a photon energy of 100 eV ~\cite{bostwick,zhou}.

\end{document}